# Synthesis of luminescent terbium-thenoyltriflouroacetone MOF nanorods for green laser application

D.Y. Medina-Velazquez[a,*], U. Caldiño[b], A. Morales-Ramirez[c,e], J. Reyes-Miranda[a], R.E. Lopez[a], R. Escudero[d], R. Ruiz-Guerrero[e], M.F. Morales Perez[a]

[a] Universidad Autónoma Metropolitana-Azcapotzalco, División de Ciencias Básicas e Ingeniería, Av. San Pablo No 180, Col, Reynosa-Tamaulipas, C.P. 02200, CDMX, Mexico
[b] Universidad Autónoma Metropolitana-Iztapalapa, Departamento de Física, P.O. Box 55-534, CDMX, 09340, Mexico
[c] Instituto Politécnico Nacional-ESIQIE, Departamento de Metalurgia y Materiales A.P. 118-431, 07051, CDMX, Mexico
[d] Universidad Nacional Autónoma de México, Instituto de Investigaciones en Materiales, A. Postal 70-360, CDMX, 04510, Mexico
[e] Instituto Politécnico Nacional, CIITEC IPN, Cerrada de Cecati S/N. Col. Santa Catarina, Azcapotzalco, C.P. 02250, CDMX, Mexico

ABSTRACT

The metalorganic frameworks (MOFs) with lanthanides ions offer great potential in the optical area because can provide properties of flexibility, low density, low-cost methods of synthesis, and insolubility in water, which give them an advantage over traditional phosphors. In this study, a thenoyltriflouroacetone ligand (TTA) with a $Tb^{3+}$ MOF was synthesized (Tb = 10 and 50% mol) and its structural and luminescent properties were analyzed. The metalorganic compound was generated in a simple one-pot reaction from terbium nitrate and 2-thenoyltrifluoroacetone precursors at room temperature. By means of FTIR, it was confirmed the presence of carbon groups, which made possible the terbium ion chelation, and also the Tb-O bonds vibration modes. $^1$HNMR results confirm that the complex with 10% mol of $Tb^{3+}$ contains three coordinates molecules of TTA and two waters molecules. The powders exhibit rod-like morphology with size about 170 nm of diameter and a length about 2 μm; the rod-like nature of powders was confirmed by SEM and TEM analyses. By XRD it was concluded that at higher terbium concentration (TTA-50Tb sample) higher the crystallite size and crystallinity, in fact the TTA-10Tb sample shows a partial-amorphous nature. By photoluminescence analyses, the $^5D_4 \rightarrow {}^7F_J$ (J = 3, 4, 5 and 6) emissions were recorded for both synthesized samples ($\lambda_{exc}$ = 376 nm). Furthermore, it was observed that the emission intensity was enhanced in a factor of 3.5 for the TTA-50Tb. The energy transfer efficiency from TTA to $Tb^{3+}$ (antenna effect) was 0.984 for TTA-10Tb and 0.993 for TTA-50Tb. Decay time analyses indicate effective lifetime of 1.45 and 1.60 m s for the samples doped at 10 and 50%, respectively, indicating that the forbidden transition rules are stronger at higher crystallinity. The integrated intensities of the $^5D_4 \rightarrow {}^7F_5$ (green at 541 nm) and $^5D_4 \rightarrow {}^7F_6$ (blue at 486 nm) emissions and their intensity ratios $I_G/I_B$ upon 376 nm excitation have been evaluated for TTA-10Tb andTTA-50Tb samples. The CIE1931 color of the MOFs excited at 376 nm attains a higher green color purity by increasing the terbium concentration. This is in concordance with the increased $I_G/I_B$ ratio up for the TTA-10Tb and TTA-50Tb samples. Thus, the TTA-50Tb sample exhibits a green color purity of 67.94% with chromaticity coordinates (0.30, 0.57), being very close to those (0.29, 0.60) of European Broadcasting Union illuminant green. This interesting feature of the TTA-50Tb sample, together with an experimental branching ratio of 61.3% for the $^5D_4 \rightarrow {}^7F_5$ green emission, highlights its capability as solid state green laser pumped by GaN (376 nm) LEDs.

## 1. Introduction

Recently, the study of luminescent compounds has grown due to the large number of areas in which these materials can be applied (optics, medicine, electronics, etc.). Special attention is located on lanthanide-doped materials because their high quantum yield, narrow spectral emission and large lifetime, which depend on the host and the crystal field around lanthanide ions.

Among the many research areas of luminescence materials, the study of rare earth doped organic binders (like β-diketones) is growing fast, since it is possible to increase considerably the emission intensity of the compounds due to the so called "antenna effect" and the charge transfer of the system of singular and triplet (T) levels (S) at internal levels of the rare earth ion [1–4]. In addition, the complexes are mechanically flexible which make them suitable for wearable electronics which can be bent or folded keeping their lighting properties [5]. From these materials, TTA is a very promising luminescent organic-material, showing excellent fluorescence properties on the basis of energy transfer from the organic ligand to the central rare earth ions [6]. These properties have been used to manufacture three-dimensional screens, white LED, fluorescent label, and even solid-state lasers [7–10]. In this regard, green laser emitting at 547 nm based on terbium trifluoroacetylacetone in liquid solution was first reported by S. Bjorklund et al. [11], and their results showed that complex displayed a great stability without losing the threshold pumping energy about 1700 J, demonstrating that the organic materials can be used for laser applications.

* Corresponding author.
*E-mail address:* dyolotzin@azc.uam.mx (D.Y. Medina-Velazquez).

Regarding to material science and engineering, materials with nano/microfiber structures are of great interest for their excellent properties and potential applications in many fields, due to their extremely high surface-to-volume ratio, tunable porosity, and the ability to control fiber composition to achieve desired performances of its properties and functionalities [12,13]. Among the various synthesis methods for organic luminescent materials, the chemical process has been considered as one of the most promising synthetic routes, due to its low cost due the room temperature synthesis conditions, high efficiency and good crystallinity of the products obtained [14].

In this work, we report the synthesis of terbium doped TTA with rod-like morphology obtained by means of an easy one pot reaction synthesis. The results indicate that higher the Tb content, higher the crystallinity and thus the light emission intensity of the powders is enhanced in a factor of 3.5. The TTA-50Tb sample when excited at 376 nm exhibit a green color purity of 67.9% with CIE coordinates (0.306, 0.574) close to the (0.29, 0.60) of European Broadcasting Union illuminant green. Besides, the branching ratio of 61.3% of the $^5D_4 \rightarrow {}^7F_5$ green emission, make its available as solid state green laser pumped by GaN (376 nm) LEDs [15].

## 2. Materials and methods

For the synthesis of TTA-Tb, all precursors: Thenoyltrifluoroacetone ($C_8H_5F_3O_2S$, 99.99%), Terbium (III) nitrate pentahydrate (Tb $(NO_3)_3 \cdot 5H_2O$, 99.99%), Sodium Bicarbonate ($NaHCO_3$, 99.99), were purchased from Sigma Aldrich. Besides, distilled water and ethanol were used as solvent and used without further purification.

In a typical synthesis, 15.5 mmol of Thenoyltrifluoroacetone were dissolved into 15 mL of ethanol and kept under vigorous stirring at room temperature for 30 min. In another flask, terbium nitrate was dissolved in 10 ml of deionized water, and from this solution an appropriate volume was dropped to the Thenoyltrifluoroacetone solution to fix the terbium composition at 10 and 50% mol (TTA- 10 Tb, and TTA-50Tb). The $NaHCO_3$ was added to the solution with the aim to adjust the pH at 7. Finally, the solution was dried at 100 °C for 24 h, obtaining a white powder. For the sample TTA-Gd, same procedure was carried out from gadolinium nitrate at 10% mol.

The TTA-Tb powders were characterized by FTIR using a Perkin Elmer 2000 model. For sample preparation, the KBr pellet technique was employed, and the measurements were recorded from 4000 to 400 $cm^{-1}$. In order to know the coordination of the MOF, the HNMR spectra was conducted on a Bruker 750 MHz spectrometer (Bruker Biospin, Rheinstetten, Germany) equipped with a 5 mm TXI cryoprobe. The $^1$HNMR spectra were referenced to the methyl signal of internal TMS. The sample of Tb $(TTA)_3(H_2O)_2$ were measured at 298.1 ± 0.1 K, without rotation and with 4 dummy scans prior to 64 scans. For sample preparation the sample was prepared in $CDCl_3$, due to its low solubility the sample was passed for 1 h by a sonicator prior to its analysis and the supernatant solvent was analyzed by $^1$H NMR.

Structural analysis of the prepared powders was carried out by XRD in reflective mode on a Bruker eco D8 ADVANCE diffractometer with Ni-filtered Cu Kα1 radiation (λ = 0.15406 nm). Diffraction angles (2θ) in the 10–60° range were analyzed at a step size of 0.02°-seg$^{-1}$. With the aim to know the powders morphology, it was used a scanning electron microscopy JEM-2200FS model operating at 80 keV, equipped with energy dispersive spectroscopy (EDS) for elemental mapping. Transmission electron micrographs were obtained on a transmission electron microscopy JEOL ARM-200F model operating at 200 keV accelerating voltage.

For luminescence analyses, excitation and emission spectra were obtained by using a Horiba Jobin-Yvon Flourolog 3–22 spectrofluorometer equipped with a 450 W ozone-free Xe lamp for the steady state mode and a pulsed Xe lamp for decay time measurements. Decay time profiles were recorded in the phosphorescence mode using a delay time of 0.01 m s after the excitation pulse (3 μs half-width) and a 10 m s sample window for emission decay of the terbium $^5D_4$ level.

## 3. Results

### *3.1. FTIR analyses*

FTIR spectroscopy was conducted in order to examine the chemical composition of synthesized powders. The spectra of the TTA- 10 Tb, and TTA-50Tb samples are depicted in Fig. 1. A broad band in the range 3600–3000 $cm^{-1}$ in the samples is attributable to the stretching vibrations of hydroxyl groups (υOH), arising from water chemically and physically adsorbed on the powders [13]. From the chemical structure of TTA, the presence of double carbon C=C is confirmed at 1688 $cm^{-1}$ [14–17]. The unreacted C-F bonds present in TTA observed at 1202 and 1142 $cm^{-1}$ are attributed to the asymmetrical and symmetrical stretching vibration of C-$CF_3$ [18,19]. Another intense sharp band is located at 722 $cm^{-1}$; which can be related to the C-H bonds present in the acetone ring [16]. For both samples, the signal located at 520 $cm^{-1}$ can be related to the Tb-O vibration band, indicating that the Terbium ion has effectively been incorporated into the complex. Besides, comparing the two samples, there is no considerable changes in the kind of absorption bands, but when the terbium concentration is increased from 10 to 50% molar, a shift to shorter wavenumber values occurred, which means a strong interaction among the non-coordinated oxygen atoms of TTA and the terbium ions [20–22]. However, there is no evidence of other Tb interactions, which could be consequence of different Tb-surroundings.

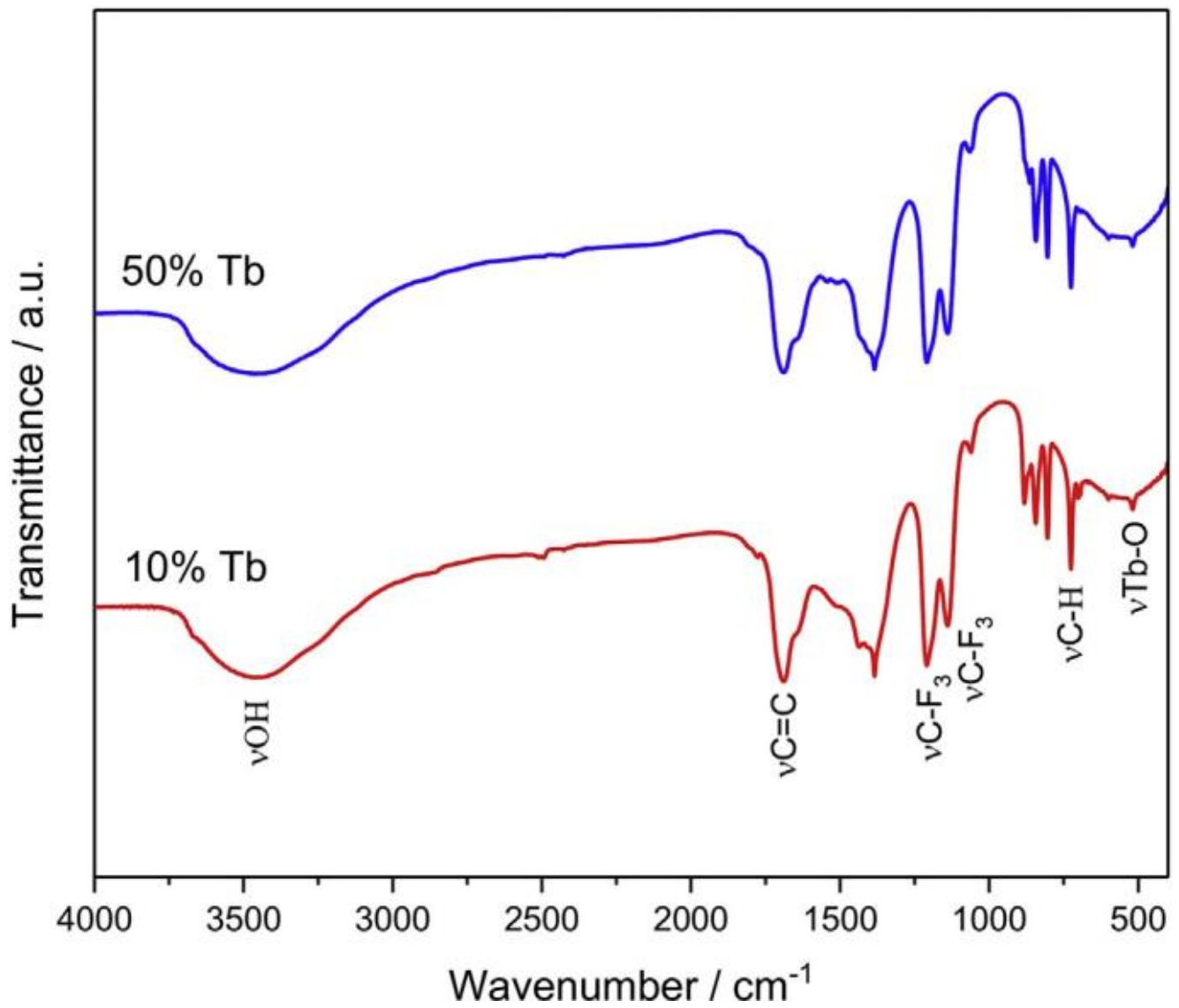

Fig. 1. FTIR spectra of the TTA- 10 Tb, and TTA-50Tb samples.

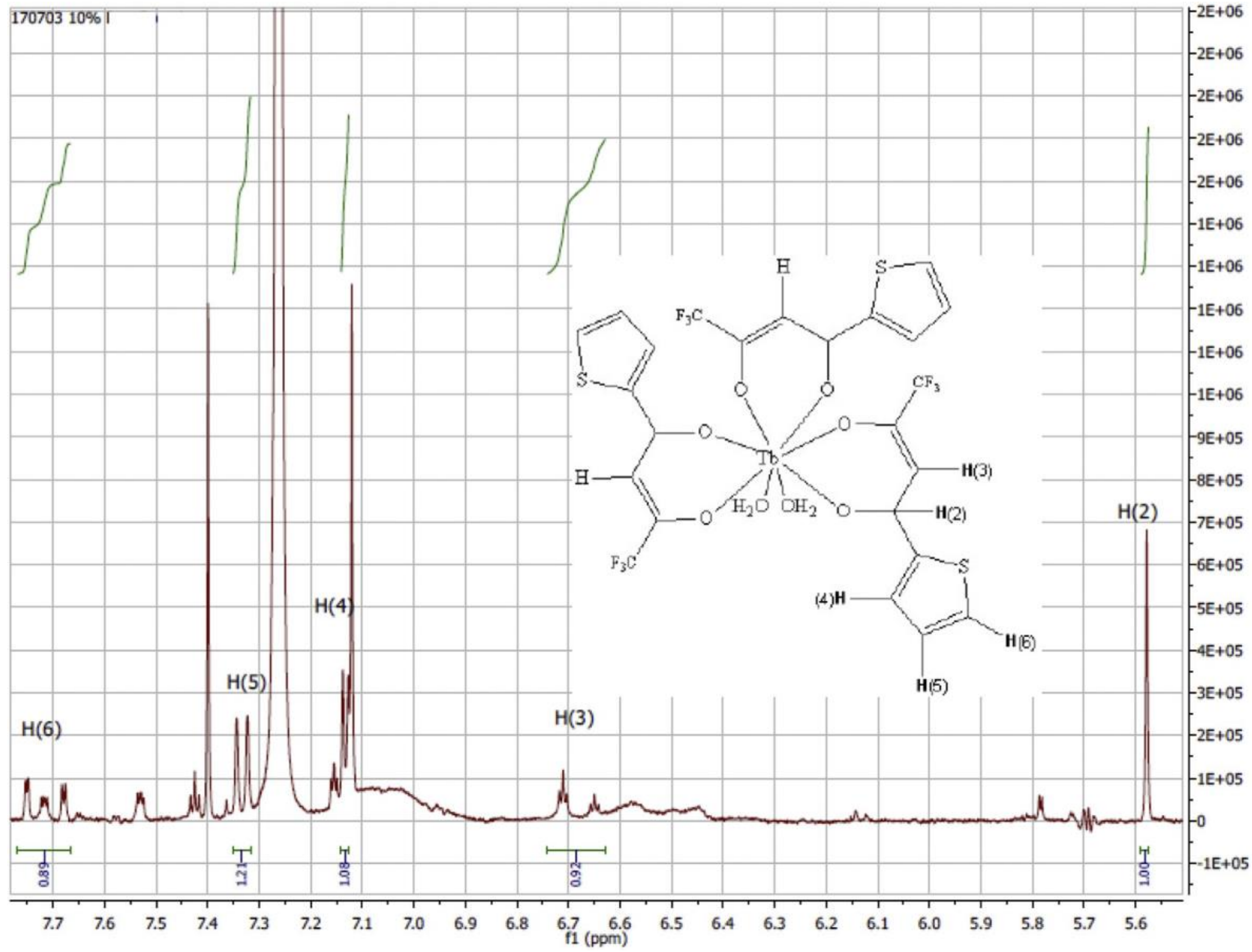

Fig. 2. Partial [1]H NMR spectra of the binary Tb(III)–TTA system measured in $CDCl_3$ with the signal assignments. The large signal corresponding to $CDCl_3$ at 7.26 (t) not deuterated, is due to the low solubility of the complex Tb $(TTA)_3(H_2O)_2$. It was decided not to eliminate it in order to observe the signals belonging to the TTA ligands.

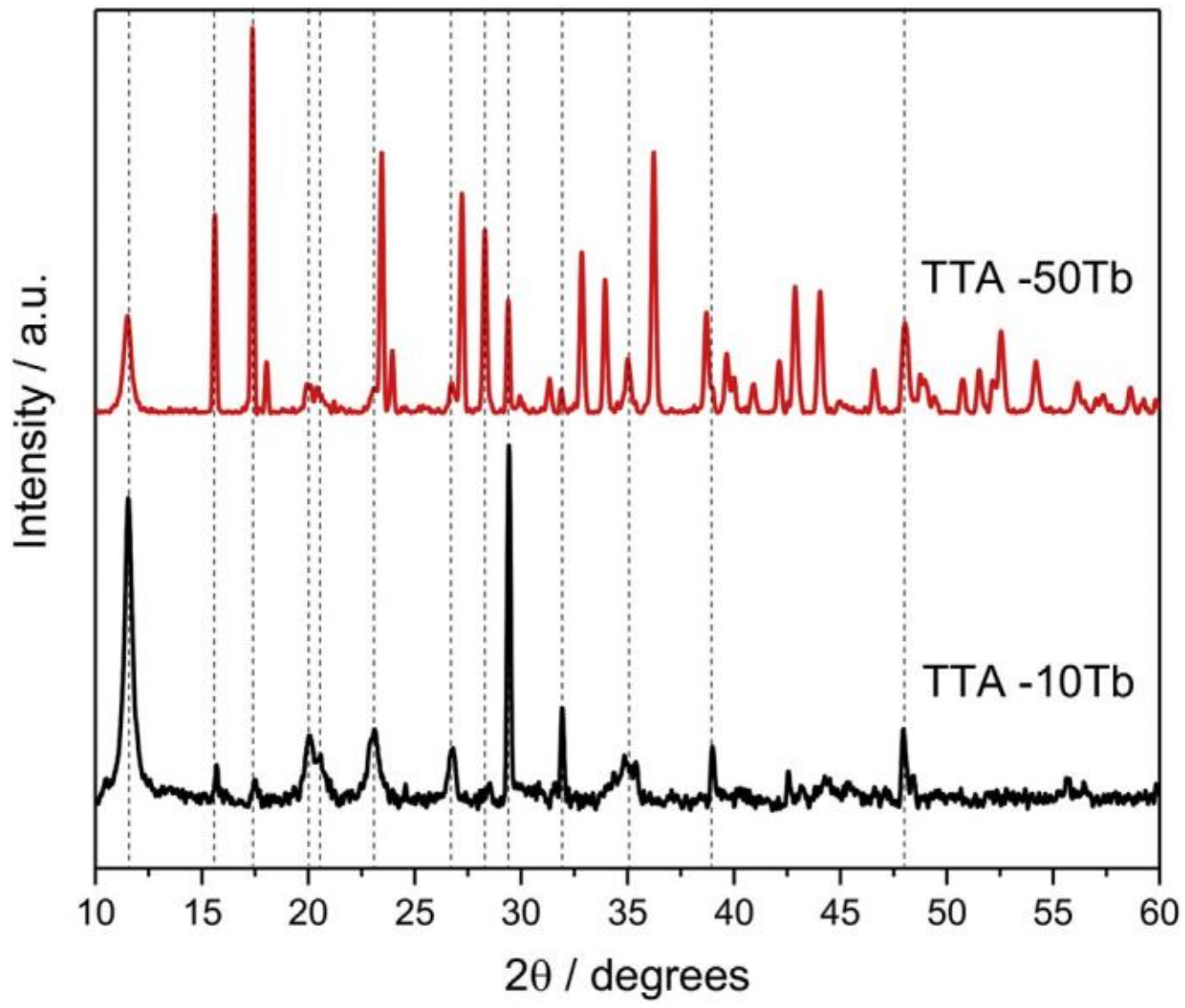

Fig. 3. XRD patterns of the TTA- 10 Tb, and TTA-50Tb samples.

### 3.2. *Stoichiometry of complex TTA-Tb*

In the TTA-Tb like in other organic compounds like TTA-TTPPO-Tb [17], the central $Tb^{3+}$ ions are coordinated to eight oxygen atoms, in this case linked with three bidentate TTA ions and two water molecules. It was confirmed by [1]H NMR spectra [23]. The sample taken from the experiment with $CDCl_3$ was Tb-TTA 10%, This choice is due to the Tb paramagnetism, so at low concentration levels it facilitates the analysis of the additional broadening bands.

The integrals of the [1]HNMR signals of coordinates ligands were used to determine stoichiometry of the complex. This integration confirmed that the complex contains three coordinates molecules of TTA and two waters molecules. The assignment of the signals corresponding to each of the protons present in the TTA ligand it's shown in the scheme of Fig. 2.

Table 1
Crystallite size computed by Scherrer equation of TTA-Tb powders.

| Peak at 2θ | Crystallite size (nm) | |
|---|---|---|
| | *TTA*-10Tb | *TTA*- 50 Tb |
| 11.5 | 19.2 | 19.6 |
| 15.7 | 38.2 | 46.84 |
| 17.5 | 29.4 | 47.34 |
| 20 | 9.5 | 22 |
| 23.2 | 13.7 | 43 |
| 29.4 | 43.3 | 49 |
| 34.9 | 6.8 | 30.3 |

### 3.3. *Structural and morphological studies*

In order to study the effect of terbium concentration on the crystal structure of TTA-Tb powders, XRD analyses were conducted and shown in Fig. 3. As observed, both samples are crystalline, but an sligthly amorphous nature is clearly evident whit the TTA- 10 Tb sample suggesting that $Tb^{3+}$ ions promote a better crystallization process. This can be due to the fact that at higher doping contents, the coherence length (the minimum size of an aggregate in out-of-plane direction) significantly decreases, and therefore the aggregates at higher Tb content would be larger [24]. In addition, the patterns show various diffraction peaks, which depend on the terbium concentration. In the first region from 10 to 30°, the same peaks appear at the two samples. Thereafter, the amount of peaks are increased in TTA- 50 Tb sample. Another interesting observation is that the intensity of the peaks varies enormously, which could indicate a morphological change depending on the $Tb^{3+}$ content. Besides, according to the Full Width at Half Maximum (FWHM) of the peaks, the TTA- 10 Tb sample shows more broad peaks, indicating a smaller crystallite size. In this regard, the crystallite size was computed by Scherrer equation:

$$D = \frac{K\lambda}{\beta cos\theta} \quad (1)$$

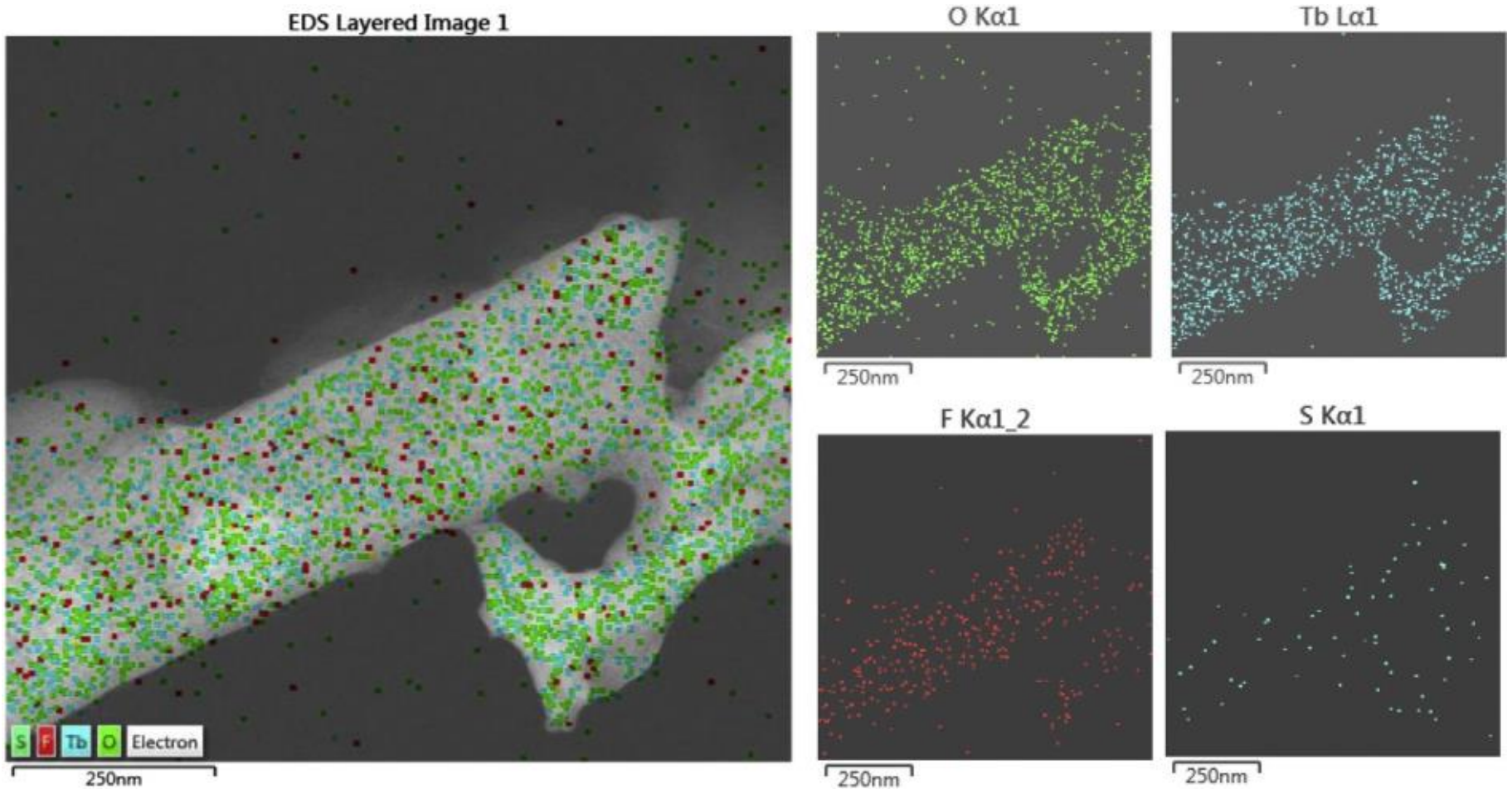

Fig. 4. Elemental mapping of TTA- 10 Tb sample.

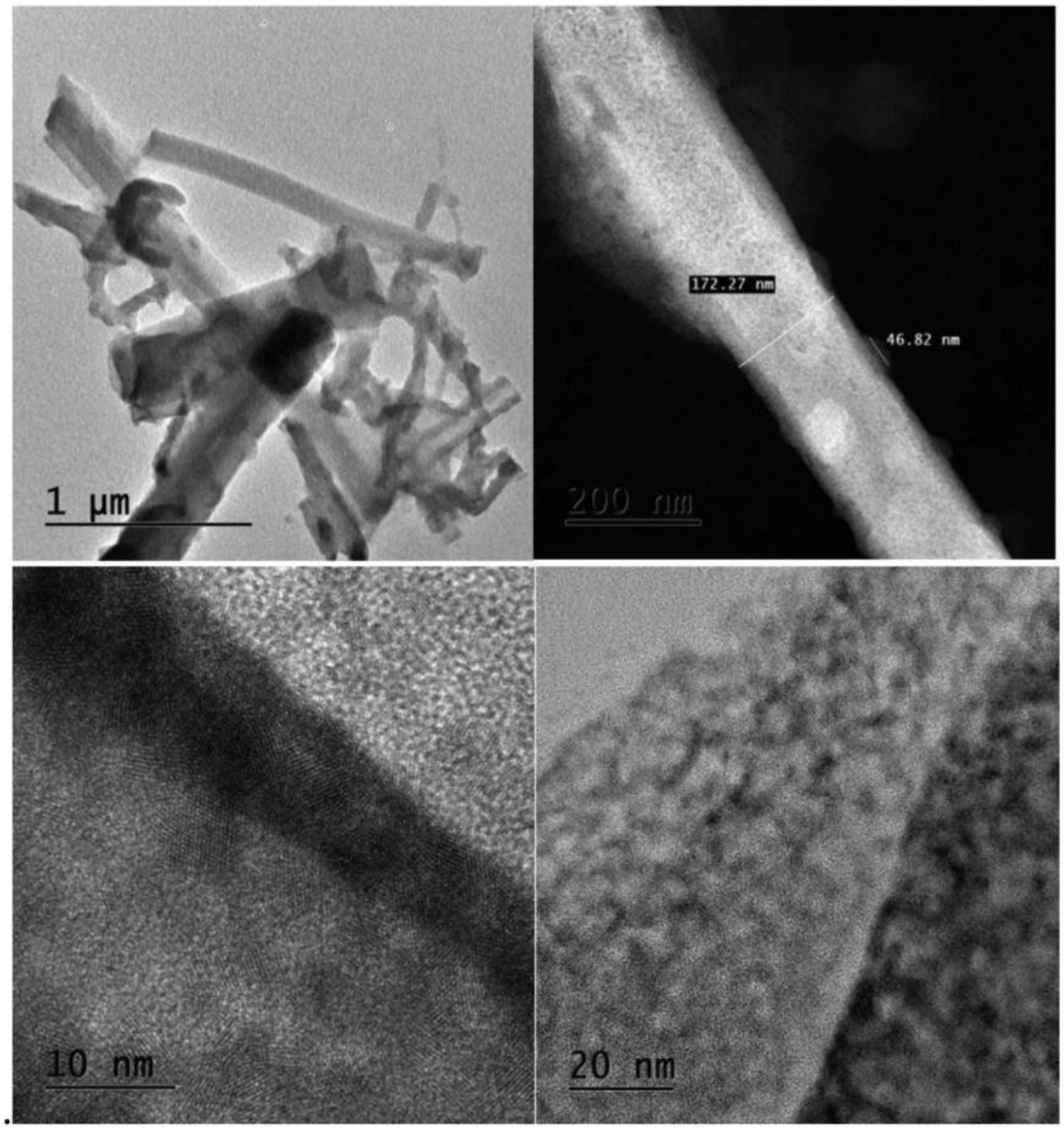

Fig. 5. TEM images of TTA-10Tb powder. The rod-like morphology is confirmed.

Where D is the crystallite size, K is a constant (0.9), $\lambda$ is the X-ray wavelength (Cu-$K_\alpha$ = 0.15418 nm), β is the full width at half maximum (FWHM) and θ is the half diffraction angle of the peak centroid. The computed crystallite size obtained from eq. (1) for both samples are

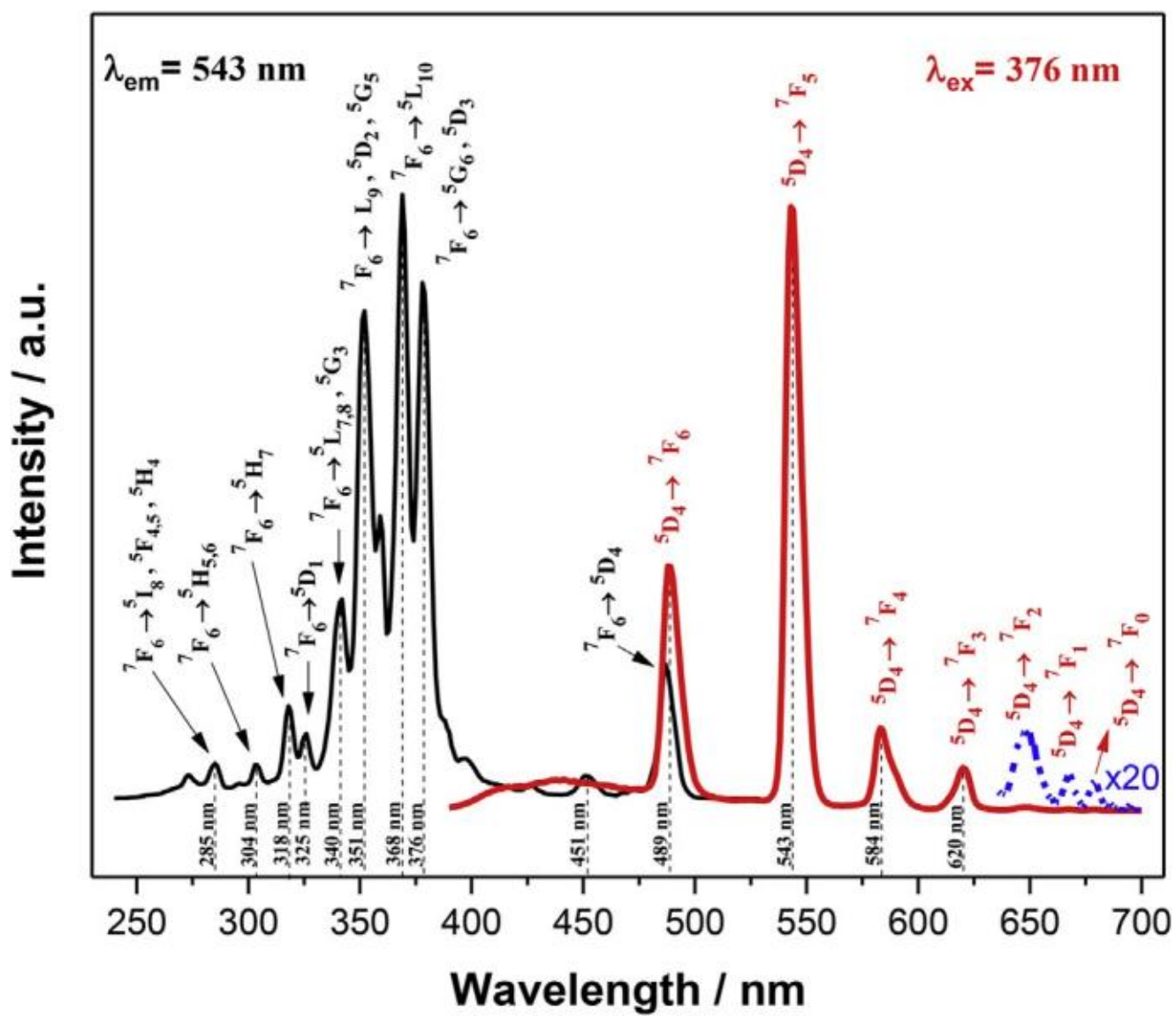

Fig. 6. Excitation and emission spectra of the TTA-10Tb sample.

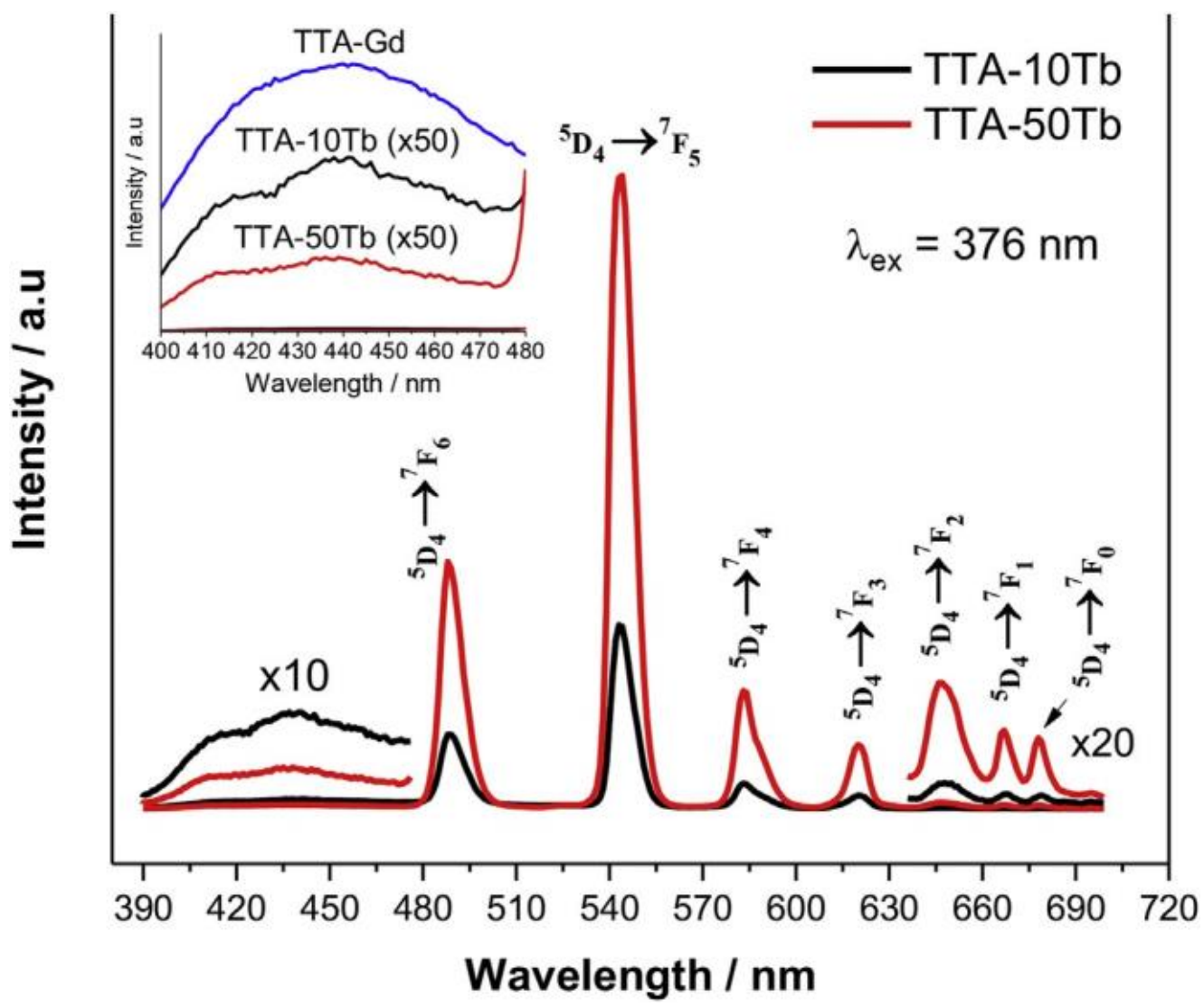

Fig. 9. Emission spectra of the TTA-10Tb and TTA-50Tb samples. Inset: Emission at 450 nm of TTA-Gd, TTA-10Tb, and TT-50Tb.

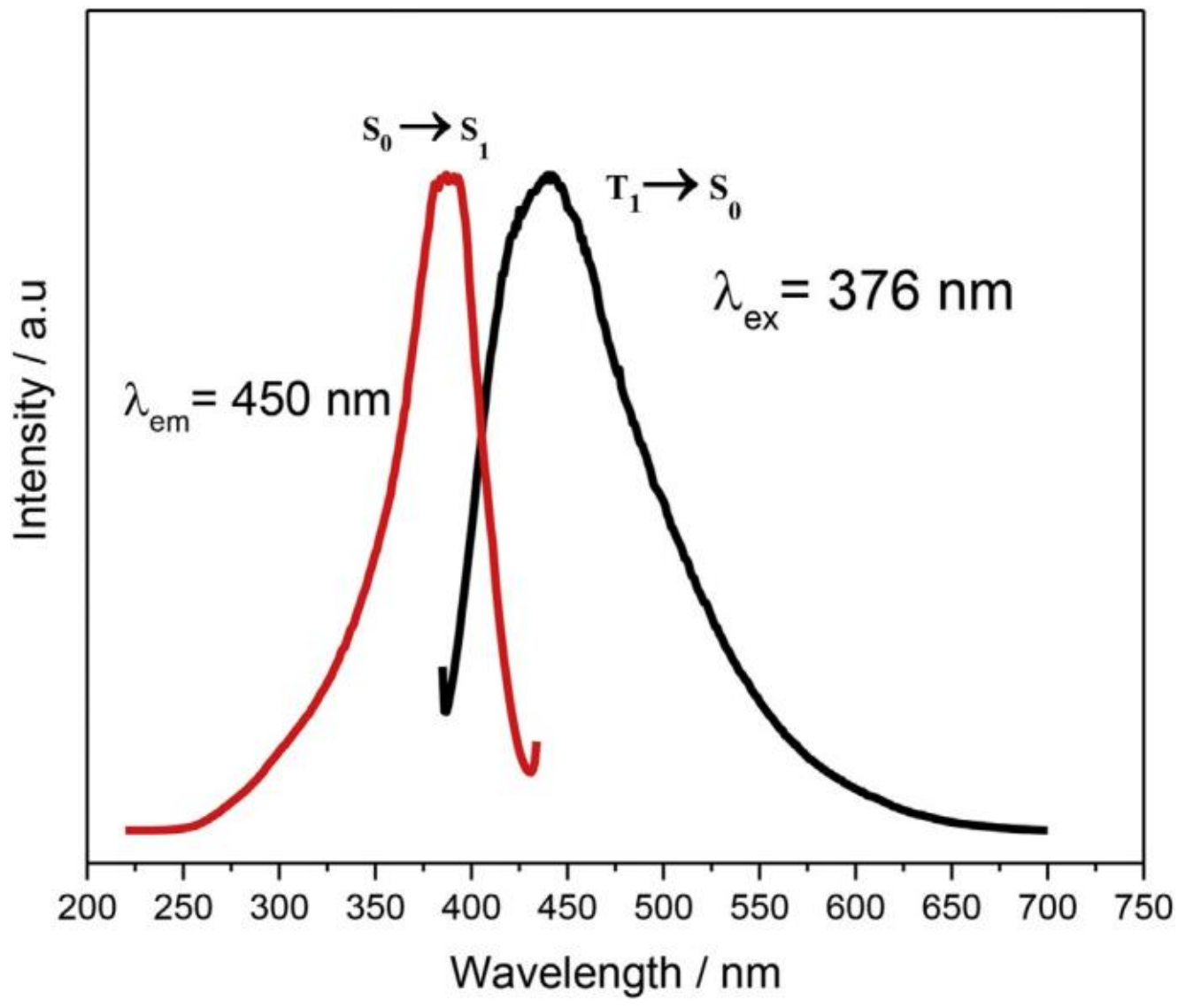

Fig. 7. Excitation and emission spectra of the TTA-Gd sample.

Table 2
Energy transfer efficiency of TTA→Tb in TTA-10Tb and TTA-50Tb samples.

| Energy transfer efficiency | | |
|---|---|---|
| Sample | $I_S$ | η |
| TTA-10Tb | 89150.47 | 0.984 |
| TTA-50Tb | 36267.00 | 0.993 |

Table 3
Branching ratios (%) of the $^5D_4 \rightarrow {}^7F_J$.

| Sample | $^5D_4 \rightarrow {}^7F_6$ | $^5D_4 \rightarrow {}^7F_5$ | $^5D_4 \rightarrow {}^7F_4$ | $^5D_4 \rightarrow {}^7F_3$ |
|---|---|---|---|---|
| TTA-10Tb | 25.5 | 61.5 | 9.0 | 4.0 |
| TTA-50Tb | 23.2 | 61.3 | 10.7 | 4.8 |

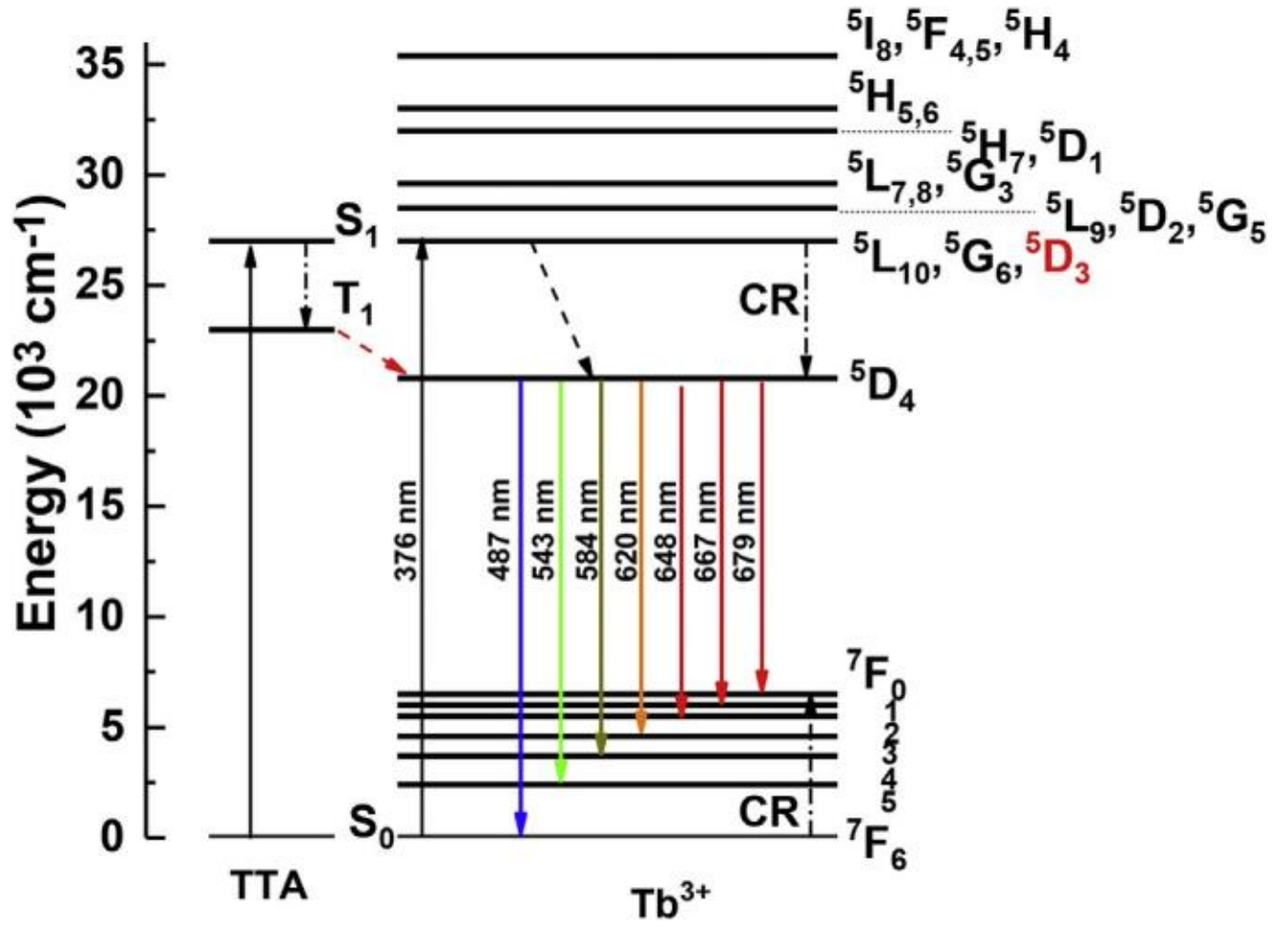

Fig. 8. Partial energy level diagram of TTA:$Tb^{3+}$ MOF.

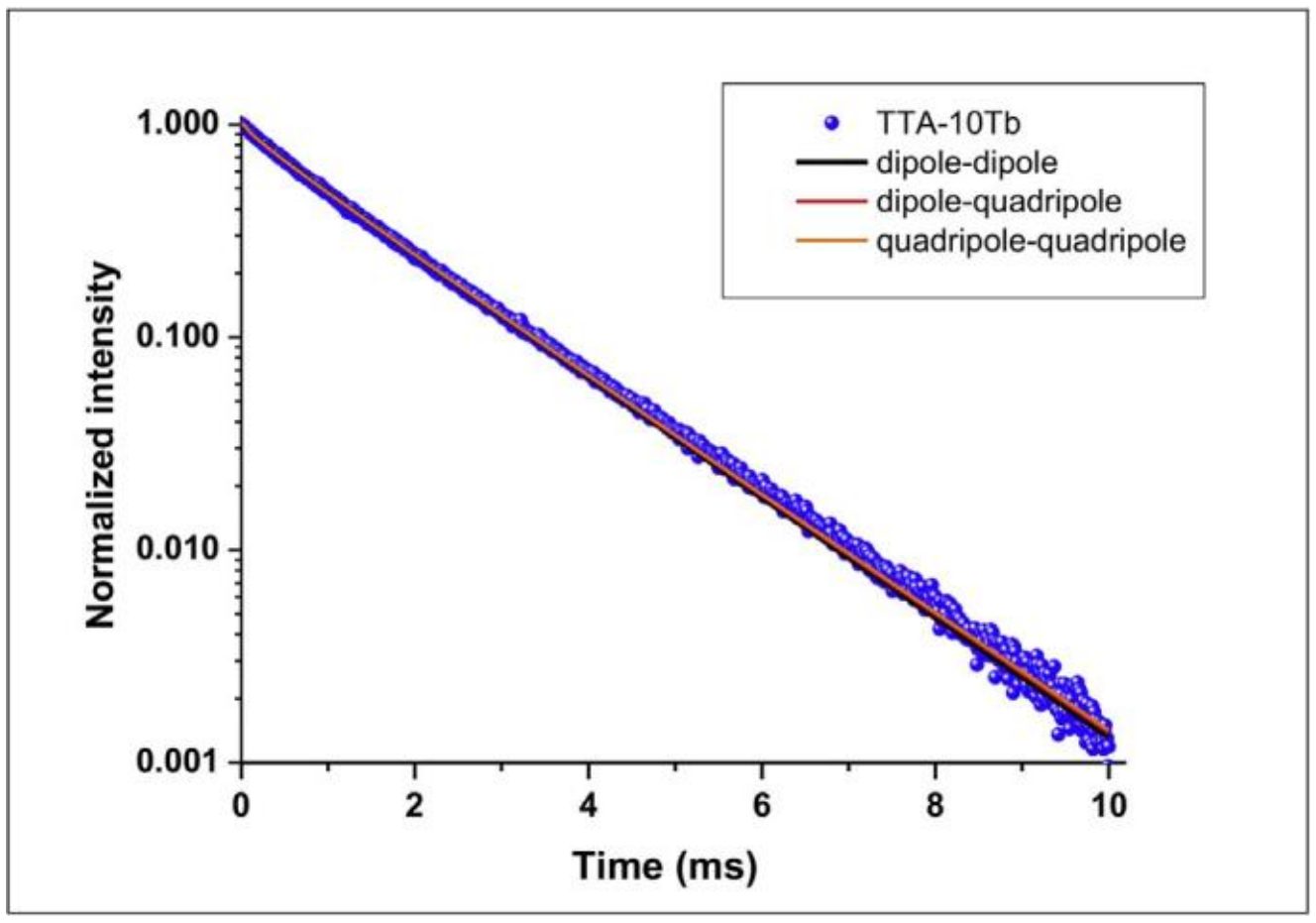

Fig. 10. $^5D_4$ level decay time profile at 543 nm of the TTA-10Tb sample excited at 376 nm.

reported in Table 1. Effectively, the crystallite size is smaller when the concentration of $Tb^{3+}$ is low (TTA-10Tb).

Fig. 4 shows an elemental mapping using energy dispersive spectroscopy (EDS) for the TTA-10Tb sample As can be observed, the distribution of terbium in the area analyzed is very homogeneous; the oxygen, fluor and sulfur are also well dispersed, and therefore there is no evidence of the possible precipitation of a different phase. The micrograph shows the rod-like morphology of the powders, which matches well with previous report [25].

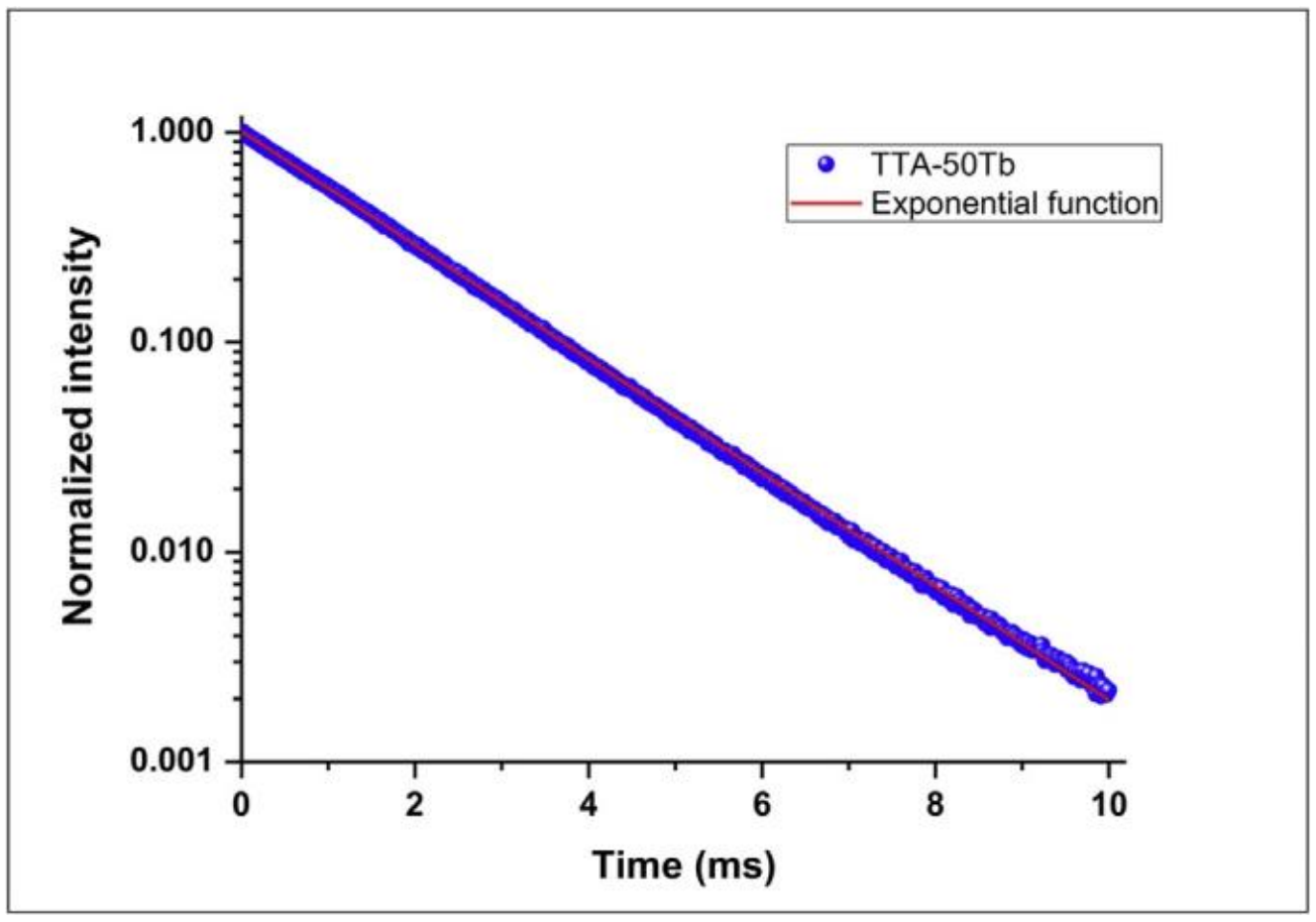

Fig. 11. $^5D_4$ level decay time profile at 543 nm of the TTA-50Tb sample excited at 376 nm.

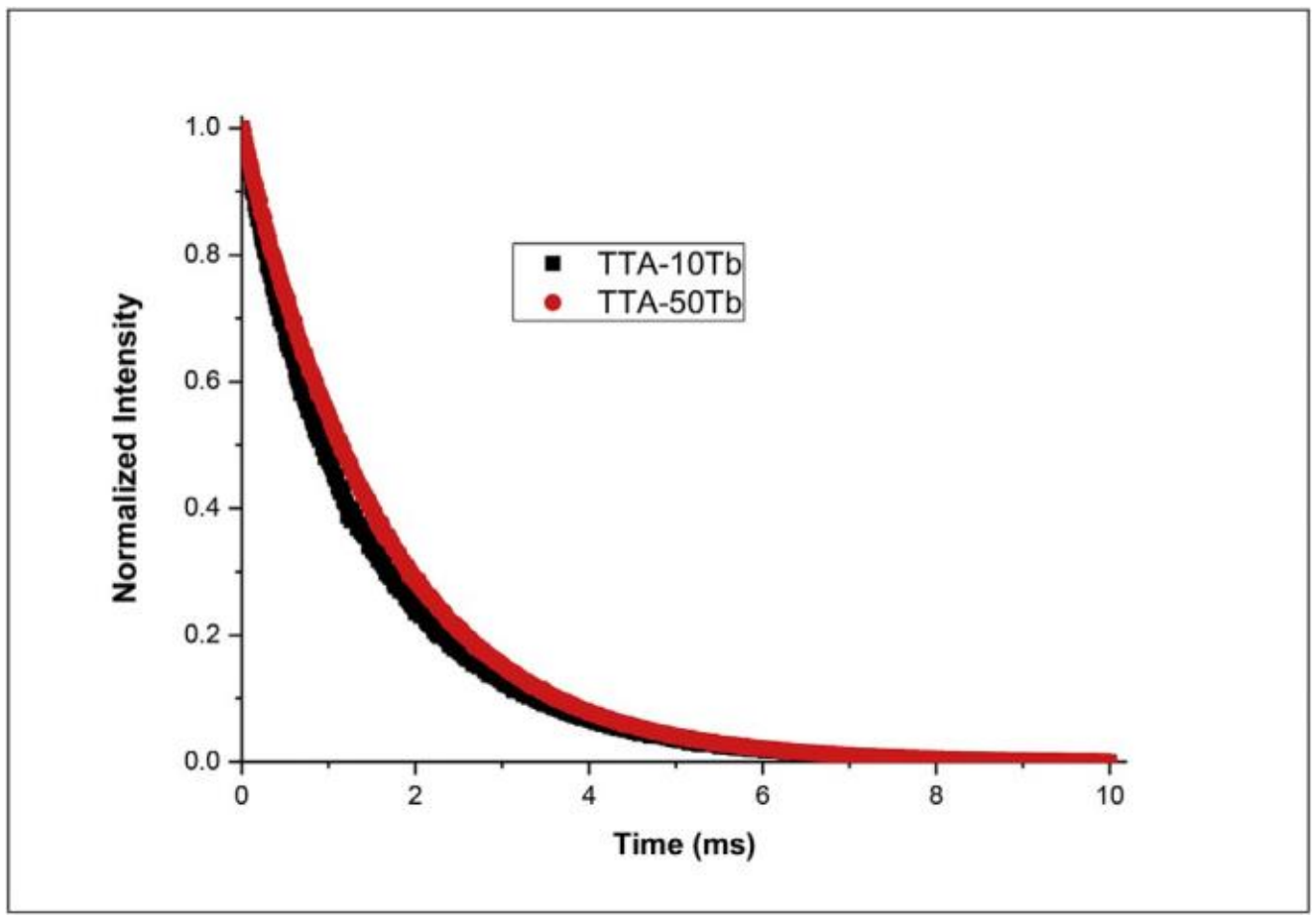

Fig. 12. $^5D_4$ level decay time profile at 543 nm of the TTA- 10 Tb and TTA-50Tb samples excited at 376 nm.

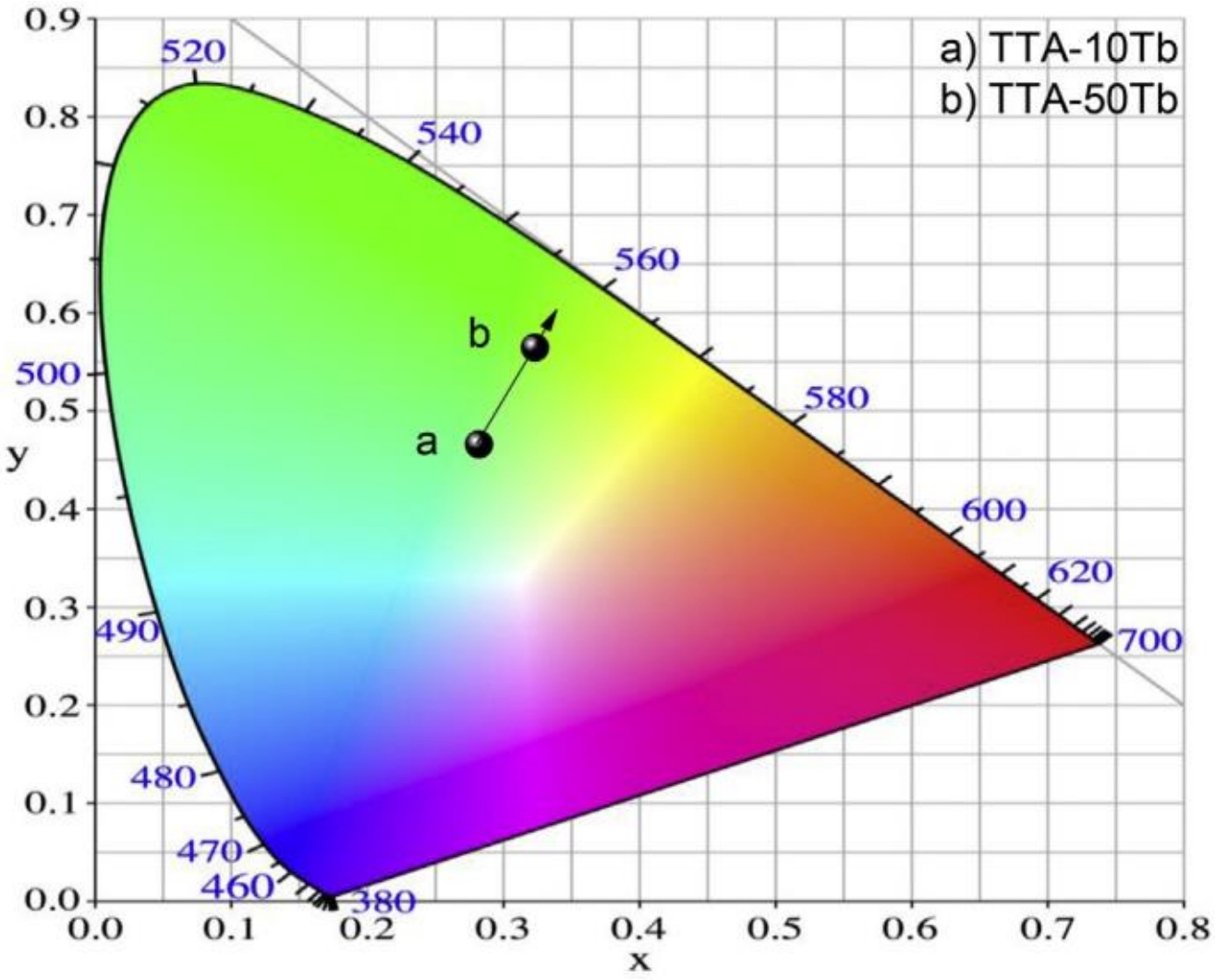

Fig. 13. Chromaticity coordinates in CIE1931 diagram. (a) TTA-10Tb and (b) TTA-50Tb.

The rod-like morphology nature of the TTA-10Tb powder was confirmed by transmission electron mycroscopy (Fig. 5). The images reveal that the rods have a diameter and length about 170 nm and 2 μm, respectively.

Table 4
CIE coordinates (*x*, *y*), CP, $I_{G\ (544)}/I_{B\ (485)}$ and lifetime of TTA-Tb powder.

| Sample | CIE coordinates | | τ (ms) | CP | $I_{G\ (544)}/I_{B\ (485)}$ |
|---|---|---|---|---|---|
| | *x* | *y* | $^5D_4$-$^7F_5$ | % | |
| TTA-10Tb | 0.265 | 0.475 | 1.34 | 53 | 2.46 |
| TTA-50Tb | 0.306 | 0.574 | 1.62 | 67.94 | 2.56 |

### *3.4. Photoluminescence*

Fig. 6 shows the excitation and emission spectra of the TTA-10Tb sample. The excitation spectrum was monitored under the $^5D_4 \rightarrow {}^7F_5$ transition of terbium at $\lambda$ = 543 nm. All the transitions from the $^7F_6$ ground level are displayed in the excitation spectrum: $^7F_6 \rightarrow {}^5I_8$, $^5F_{4,5}$, $^5H_4$ (285 nm); $^7F_6 \rightarrow {}^5H_{5,6}$ (304 nm), $^7F_6 \rightarrow {}^5H_7$, (318 nm); $^7F_6 \rightarrow {}^5D_1$, (325 nm); $^7F_6 \rightarrow {}^5L_{7,8}$, $^5G_3$ (340 nm); $^7F_6 \rightarrow {}^5L_9$, $^5D_2$, $^5G_5$ (351 nm); $^7F_6 \rightarrow {}^5L_{10}$ (368 nm); $^7F_6 \rightarrow {}^5G_6$, $^5D_3$ (376 nm); and $^7F_6 \rightarrow {}^5D_4$ (487 nm) [26–28]. On the other hand, the emission spectrum was recorded through the $^5D_3$ excitation level at $\lambda$ = 376 nm. The spectrum shows the transition from the $^5D_4$ excitation level to the $^7F_J$ multiplets: $^5D_4 \rightarrow {}^7F_6$ (489 nm); $^5D_4 \rightarrow {}^7F_5$ (543 nm); $^5D_4 \rightarrow {}^7F_4$ (584 nm); $^5D_4 \rightarrow {}^7F_3$ (620 nm). Interestingly, there is no evidence of blue emission of terbium coming from the $^5D_3$ level. However, a broad band in the blue region from 390 to 480 nm is observed, which could be attributed to a possible emission of the TTA host. In order to confirm it, a sample of TTA-Gd was synthesized because $Gd^{3+}$ ions does not interact with ultraviolet light and therefore the observed luminescence spectrum is attributed to the TTA host. Fig. 7 shows the excitation and emission spectra of TTA-Gd. As expected, the TTA host absorbs light at 370 nm, close to the $^5D_4 \rightarrow {}^7F_5$ transition of terbium at 376 nm, and emission spectrum shows a broad blue emission centered at 450 nm, which confirms that emission showed in TTA-10Tb and TTA-50 Tbat this wavelength is related to the TTA host. Fig. 8, portrays a partial energy level diagram showing the $^5D_4$ level emissions and cross relaxation $^7F_6$, $^5D_4 \rightarrow {}^7F_0$, $^5D_3$ e of $Tb^{3+}$ ions in TTA-10Tb sample excited at 376 nm, as well the excitation of the TTA at 370 nm ($S_o \rightarrow S_1$), no radiative relaxation $S_1 \rightarrow T_1$ in TTA and energy transfer $T_1 \rightarrow {}^5D_4$ from the ligand (TTA) in order to show that $Tb^{3+}$ ions were excited by directly excitation $^7F_6 \rightarrow {}^5D_3$ and via ligand TTA (antenna effect) simultaneously [29,30].

The emission spectra of the TTA-Gd, TTA-10Tb and TTA-50Tb powders excited at $\lambda$ = 376 nm are shown in Fig. 9. As it can be observed, the emission from TTA ligand at 450 nm dismiss with higher Tb content, and the emission intensity of terbium ions from the $^5D_4$ level is enhanced when the terbium concentration increases; in fact, the intensity is enhanced 3.5 times with the higher $Tb^{3+}$ concentration without luminescence quenching. Furthermore, the energy transfer efficiency can also be measured from the intensities of the sensitizer emission in the presence ($I_s$) and absence ($I_s^0$) of the activator ($Tb^{3+}$), through the following expression [31]. Table 2 shown the energy transfer efficiency for the TTA-10Tb and TTA-50Tb as it shown the energy transfer (antenna effect) increase with higher $Tb^{3+}$ concentration.

$$n = 1 - \frac{I_s}{I_s^0} \qquad (2)$$

The branching ratios (%) of the $^5D_4 \rightarrow {}^7F_J$ transitions are listed in Table 3. The transition $^5D_4 \rightarrow {}^7F_5$ displays values higher than 60%, which makes the TTA-Tb a promising material for green laser pumped by GaN (376 nm) LEDs [32].

Fig. 10 shows the $^5D_4$ level temporal decay at 543 nm of the TTA-10Tb sample excited at 376 nm, which is non-exponential. It was well fitted through the Inokuti Hirayama model for S = 10 [33]. Therefore, an electric quadrupole-quadrupole interaction might be the dominant mechanism in the cross-relaxation energy transfer occurring between $Tb^{3+}$ ions, since the direct energy transfer donor to acceptor parameter $\gamma_6$ was found to be 0.218. On other hand, as it is shown in Fig. 11 the $^5D_4$ level decay time profile at 543 nm of the TTA-50Tb sample was well fitted through a simple exponential function. This confirms that a higher crystallinity in TTA-50Tb sample leads to a higher dissolution of $Tb^{3+}$ ions in the host. It is well known that non-exponential luminescence decays arise rather from ions forming aggregates. The $^5D_4$ level decay time profiles of both samples (TTA-10Tb and TTA-50Tb) are shown in Fig. 11. The lifetime values of the both samples were obtained from the time at which the decay intensity has decayed $e^{-1}$ (0.368) of its initial value. The decay time decreases from 1.62 ms (TTA-50Tb sample) to 1.34 ms (TTA-10Tb sample) indicating that the forbidden transition rules are stronger with higher crystallinity (see Fig. 12).

Fig. 13 shows the CIE chromaticity coordinates for the TTA-10Tb and TTA-50Tb samples. As observed, it is attained a higher green color purity by increasing the $Tb^{3+}$ concentration. The values of the computed CIE coordinates, CP (color purity) and $I_G/I_B$ ratios are listed in Table 4. The increased $I_G/I_B$ ratio up in the TTA-50Tb is consistent with its higher green color purity.

The color purity of particular dominant color in a source is the weighted average of the ($x_s$, $y_s$) sample emission color and ($x_d$, $y_d$) dominant wavelength coordinates relatives to the ($x_i$, $y_i$) illuminant coordinates. Thus, the color purity (CP) compared to the CIE1931 standard source C illuminant with ($x_i$ = 0.3101, $y_i$ = 0.3162) coordinates is given by the expression [34,35]:

$$CP = \sqrt{\frac{(x_s - x_i)^2 + (y_s - y_i)^2}{(x_d - x_i)^2 + (y_d - y_i)^2}} \times 100\,\% \tag{3}$$

Thus, the green color purity obtained from Eq. (2) increases from 53% (TTA-10Tb sample) to 67.9% (TTA-50Tb sample), which confirm that the color purity increases with terbium concentration. All the optical characteristics are listed in Table 4.

## 4. Conclusions

Terbium doped TTA powders were synthesized by one pot chemical reaction at room temperature. The $Tb^{3+}$ concentration played an important role in the structural and optical properties. By XRD it was demonstrated that the higher Tb content the higher crystallinity of the powders, so that the terbium content contributed to a bigger crystallite size. The synthesis parameters promoted the formation of rod-like particles with size about 170 nm of diameter and a length about 2 µm for the TTA-10Tb sample. By photoluminescence analyses it was only observed the $^5D_4 \rightarrow {}^7F_J$ (J = 3, 4, 5 and 6) emissions; no emissions from the $^5D_3$ level was recorded on the spectra, which confirms an effective cross relaxation energy transfer between $Tb^{3+}$ ions. A TTA-Gd sample was synthesized in order to confirm the antenna effect of the TTA molecule, showing a strong bluish emission which tends to diminish with the incorporation of Tb ions. The energy transfer efficiency of Tb ions were obtained of 0.984 and 0.993 for TTA-10Tb and TTA-50Tb. Lifetimes of 1.34 and 1.62 ms were measured for the TTA-10Tb and TTA-50Tb samples, respectively indicating that the forbidden transition rules are stronger with higher crystallinity. The Inokuti-Hirayama fitting model in the $^5D_4$ level decay of the TTA-10Tb sample suggests a cross relaxation energy transfer between $Tb^{3+}$ ions through a quadrupole-quadrupole interaction. The CIE1931 color of the MOFs excited at 376 nm attains a higher green color purity by increasing the terbium concentration. This fact is consistent with the increased $I_G/I_B$ ratio up for the samples from TTA-10Tb to TTA-50Tb. Thus, the TTA-50Tb sample exhibits a green color purity of 67.94% with chromaticity co- ordinates (0.31, 0.57), being very close to those (0.29, 0.60) of European Broadcasting Union illuminant green. This interesting feature of the TTA-50Tb sample, together with an experimental branching ratio of 61.3% of the $^5D_4 \rightarrow {}^7F_5$ green emission, highlights its capability as solid state green laser pumped by GaN (376 nm) LEDs.

## Acknowledgments

This work was supported by CONACYT project 254280. R. E. Lopez-Romero thanks to CONACyT-Mexico for a scholarship, The authors wish to acknowledge the technical assistance of Ing. Juan Carlos Orozco Pliego and M. C. Vıctor H. Colin also to Eng. Oscar Francisco Rivera Dominguez, Ms.T Maribel Pacheco Ramos and Dr. Veronica Melo Mendoza for their help.